# Gravity-driven filamentary flow in molecular clouds

Raúl Naranjo-Romero⋆, Enrique Vázquez-Semadeni†, Robert M. Loughnane‡
*Instituto de Radioastronomía y Astrofísica,*
*Universidad Nacional Autónoma de México, Apdo. Postal 3-72, Morelia, Michoacán, 58089, México*

**ABSTRACT**
We present **a numerical study of the gravity-driven filamentary flow that arises in the presence of elongated perturbations embedded in a globally gravitationally unstable medium. We perform** idealized simulations of **the** gravitational collapse of a moderate filamentary perturbation with a central enhancement (a core) embedded in **either** a uniform **or** a stratified background. Both simulations maintain the filamentary structure during the collapse, developing a hierarchical accretion flow from the cloud to the filament, and from the filament to the core. The flow changes direction smoothly at every step of the hierarchy, with no density divergence nor a shock developing at the filament's axis during the prestellar evolution. The flow drives accretion onto the central core and drains material from the filament, slowing down the growth of the latter. As a consequence, the ratio of the central density of the core to the filament density increases in time, diverging at the time of singularity formation in the core. The stratified simulation best matches the observed Plummer-like radial column density profiles of filaments, while the uniform simulation does not produce a flat central density profile**, supporting** suggestions that MCs may be preferentially flattened structures. We examine the possibility that the filamentary flow might approach a quasi-stationary regime in which the radial accretion onto the filament is balanced by the longitudinal accretion onto the core. A simple argument suggests that such a stationary state may be an attractor for the system. **Nevertheless,** our simulations **only marginally approaches** it during the prestellar stage.

**Key words:** Gravitation – hydrodynamics – ISM: structure – ISM: clouds – ISM: evolution – Stars: formation

## 1 INTRODUCTION

### 1.1 Molecular Cloud substructure and star formation

Although observations over many decades have shown that molecular clouds (MCs) are the nurseries of star formation, the process through which MCs evolve to form stars has undergone major transformations in recent years. In particular, the existence of filamentary structure in MCs has been known for several decades (e.g., Bally et al. 1987; Nagahama et al. 1998; Myers 2009). Improved observations, especially with the *Herschel* Observatory (see, e.g., the review by André et al. 2014). In particular, as discussed by André et al. (2014), the *Herschel* continuum observations and surveys have revealed that filamentary structures are ubiquitous in MCs, that they contain the majority of the prestellar cores in MCs, and that the formation of filaments appears to precede star formation.

In addition, numerical simulations at the scale of a few ×100 **pc describing the formation of giant molecular clouds (GMCs) by generic compressions in the warm diffuse medium, and following their evolution** up to their star-forming stage, systematically develop filaments which accrete from the cloud scale and in turn accrete onto the cores they contain. This occurs regardless of whether the simulations include only hydrodynamics, cooling and self-gravity (e.g., Vázquez-Semadeni et al. 2007; Heitsch et al. 2009b; Gómez & Vázquez-Semadeni 2014; Carroll-Nellenback et al. 2014; Heiner et al. 2015; Camacho et al. 2016, 2020), include also some form of stellar feedback (Smith et al. 2011; Colín et al. 2013), include magnetic fields but no feedback (Heitsch et al. 2009a; Fogerty et al. 2016, 2017; Zamora-Avilés et al. 2018) or include both of them (Zamora-Avilés et al. 2019). In general, the role of various ingredients, such as shearing, inhomogeneities, alignment between the flow and the magnetic field, or the strength of the magnetic field is to slow down the gravity-driven flow and the subsequent formation of dense cores. Nevertheless, the main pattern of the flow is the same, developing large-scale collapse along filamentary structures, through the "edge effect" mechanism outlined by Burkert & Hartmann (2004), itself a manifestation of the amplification of anisotropies produced by pressureless collapse (Lin et al. 1965).

It is important to emphasize that, in all of these large-scale simulations, it is the above process, rather than *strongly* supersonic turbulence,[1] that dominates the formation of filamentary structures. In addition, *external* turbulence driving by supernova explosions is generally not considered to prevent this regime of flow (Iffrig & Hennebelle 2015; Ibáñez-Mejía et al. 2017, although see Padoan et al. 2016 for a differing view).

---

⋆ E-mail: rnaranjoromero@gmail.com
† E-mail: e.vazquez@irya.unam.mx
‡ E-mail: loughnane.robert@gmail.com

One way to interpret this multi-stage collapse process is in terms of the recently proposed global, hierarchical collapse (GHC), in which small-scale collapses occur within larger-scale ones and are shifted in time (collapses within collapses; Vázquez-Semadeni et al. 2019). Moreover, the larger-scale collapses produce filamentary structures because they occur in a nearly pressureless form, due to the large number of Jeans masses they contain. As a consequence, they collapse fastest along their shortest dimension, in which anisotropies are amplified, thus forming sheets and filaments (*e.g.*, Lin et al. 1965; Gómez & Vázquez-Semadeni 2014). Hence, filaments correspond to the intermediate-scale collapse, between the cloud and the core scales. It is noteworthy that the latter authors argue that the collapse of clouds proceeds from the cloud at large to sheet-like objects, then to filaments, and finally to clumps, so it is possible that the filaments tend to be embedded in sheet-like clouds. A more detailed and comprehensive description of the GHC scenario has recently been described by Vázquez-Semadeni et al. (2019), and put in the context of the lifecycle of Giant Molecular Clouds in the Galaxy by Chevance et al. (2020).

Observations from the *Herschel* Gould belt survey (*e.g.*, Arzoumanian et al. 2011; Palmeirim et al. 2013) have also found that the filaments have radial column density profiles falling off as $r^{-1.5}$ to $r^{-2.5}$ at large radii, and flattened profiles at small radii, with typical widths $\sim 0.1$ pc, although the universality of this width has been questioned by Panopoulou et al. (2017), who argued that it is an artifact of sampling a truncated power-law distribution with uncertainties.

Filaments appear to be highly dynamic entities. Rivera-Ingraham et al. (2017) have shown evidence that the filaments evolve from a subcritical (i.e., stable; Inutsuka & Miyama 1992) to a supercritical (*i.e.*, unstable) regime by accretion of material from their environment according to recent observations of nearby ($d < 500$ pc) filaments (Rivera-Ingraham et al. 2016). They found that self-gravitating filaments in dense environments ($A_v \sim 3$, $N_{H_2} \sim 2.9 \times 10^{20}$ cm$^{-2}$) can become supercritical on timescales of $\sim 1$ Myr, and suggested that filaments evolve in tandem with their environment. Also, Arzoumanian et al. (2013) found that thermally subcritical filaments have transonic velocity dispersions independent of their column density, while thermally supercritical filaments have higher velocity dispersions scaling roughly as the square root of the column density. They suggest that the higher velocity dispersions of supercritical filaments may not directly arise from supersonic interstellar turbulence, but instead may be driven by gravitational contraction/accretion. Finally, a number of molecular-line studies of filaments have suggested that there is a net gas flow *along* the filaments, perhaps feeding the hubs and clumps as a consequence of the global collapse of the filaments (e.g., Schneider et al. 2010; Peretto et al. 2013; Kirk et al. 2013).

Many existing analytical models (e.g., Ostriker 1964; Inutsuka & Miyama 1992; Fischera & Martin 2012) consider hydrostatic equilibrium, while others do consider accretion (e.g., Heitsch 2013a,b; Hennebelle & André 2013). However, to our knowledge, an analytical model for the fundamental underlying flow regime, in the spirit of, for example, the classic study by Bondi (1952) on spherically symmetric accretion, has not been performed for the case of cylindrical geometry, including the presence of a central spherical core within the cylinder. Our study represents a numerical attempt at achieving a similar goal: to characterize the properties of the gravitationally driven accretion flow into a central core funneled by a cylindrical filament, as was observed in the molecular cloud evolution simulation of Gómez & Vázquez-Semadeni (2014). For this reason, we deliberately omit any complicating factors such as turbulence or magnetic fields, and limit our study to just the flow generated by self-gravity in the presence of a cylindrical plus a spherical perturbation.

Our approach is thus different from that of most existing simulations of filamentary structures, which have been aimed at investigating the effects of various physical processes, such as turbulence and magnetic fields, on isolated filaments. For example, Seifried & Walch (2015) have investigated the stabilization against radial collapse of filaments by turbulence and magnetic fields; Clarke et al. (2016, 2017, 2020) have investigated the fragmentation of filamentary structures and the growth rates of the perturbations; Heigl et al. (2016) have furthermore investigated the consumption of gas in the filaments due to accretion onto the cores, but neglected accretion onto the filament, while Heigl et al. (2018) and Heigl et al. (2020) have investigated the accretion-driven generation of turbulence in the filaments, in the absence and presence of self-gravity, respectively, and Gritschneder et al. (2017) have investigated the triggering of collapse of marginally stable filaments by bending modes on the filaments. Instead, in this paper, we choose to rid the problem of all complicating agents, such as magnetic fields and turbulence, and to reduce it exclusively to an investigation of the flow regime generated by gravitational contraction in the presence of a filamentary perturbation with a central density enhancement.

Our approach also differs from most other studies in the sense that the system we study is *globally out of equilibrium*, allowing for global collapse in the numerical box (except at its boundaries due to the boundary conditions; cf. Sec. 4.3.2). However, it seems to approach a quasi-stationary flow regime, in which the longitudinal mass flow along the filament onto the central core approximately compensates the accretion flow from the cloud onto the filament. From this point of view, this work constitutes an extension of our previous study of the collapse of an idealized spherical core embedded in a uniform, unstable background medium (Naranjo-Romero et al. 2015, hereafter Paper I), by adding a filamentary perturbation that triggers non-spherical collapse motions. We investigate the fundamental underlying flow pattern in a filamentary structure that is part of the collapse flow from the clump scale (a few parsecs) down to the core scale (a few times 0.01 pc), and discuss whether or not this regime is capable of explaining some of the observed structural and kinematic features of MC filaments. In Sec. 2, we first present the numerical simulations, and in Sec. 3, we present the results concerning the flow pattern, the approximation to a stationary state, the evolution of structural features, and a comparison with observations. Finally, in Sec. 4, we discuss and summarize our main results.

---

[1] *Moderately* supersonic turbulence is indeed present, and provides the density fluctuations that can grow by gravity as the global collapse proceeds (Vázquez-Semadeni et al. 2019), but does not directly induce the collapse of the compressed regions (Clark & Bonnell 2005).

## 2 THE SIMULATIONS

We use a spectral, fixed mesh numerical code (Léorat et al. 1990; Vázquez-Semadeni et al. 2010) to perform two numerical simulations of hierarchically collapsing density perturbations initially at rest. These simulations consist of a spherical density enhancement with a Gaussian radial profile ("the core") embedded in a cylindrical perturbation, also with a radial density profile ("the filament"), which in turn is immersed in a uniform density background ("the cloud"). One simulation includes a stratification representing the possibility that the clouds may be flattened (e.g., Beaumont & Williams 2010; Veena et al. 2017; Kusune et al. 2019). As in Paper I, we restrict our study to the prestellar stage of the evolution because our code does not include a prescription for the creation of sink particles.

We consider an isothermal gas with mean number density $n = 100\,\mathrm{cm}^{-3}$, mean particle weight $\mu = 2.36$, temperature $T = 11.4$ K, and isothermal sound speed $c_\mathrm{s} = 0.2\,\mathrm{km\,s^{-1}}$ in a numerical box of size $L_\mathrm{box} \approx 7.1$ pc with periodic boundaries **for all fields, including the gravitational potential**, and 512 cells per dimension, for a fixed resolution of 0.014 pc. The Jeans length is $L_\mathrm{J} = 2.24$ pc, implying a Jeans mass $M_\mathrm{J} = (4\pi/3)\langle\rho\rangle(L_\mathrm{J}/2)^3 = 34.21 M_\odot$. The mass in the numerical box is $M \approx 2070\ M_\odot = 60.39\ M_\mathrm{J}$.

For the density background (*i.e.*, "the cloud"), we have considered two different configurations, corresponding to each of our two simulations. The first has a uniform density background, and therefore has *axial* symmetry, so it is labeled RunA. The second is *stratified* in the $y$ direction, and is thus labeled RunS (see Fig. 1). RunS therefore represents the case of a filament embedded in a sheet-like cloud, which is the outcome of cloud formation by the convergence of oppositely-directed gas streams in the warm, diffuse atomic gas (e.g., Vázquez-Semadeni et al. 2006; Heitsch et al. 2008; Heitsch 2013b; Wareing et al. 2019). For convenience, in this run we will refer to the central plane perpendicular to the direction of stratification as the *dense plane*. We will discuss the variation of some physical quantities in three directions: one *perpendicular* to the dense plane, one *parallel* to it (i.e., on the plane), but perpendicular to the filament, and another one on the plane but running along the filament, to which we will refer as the *longitudinal* direction. In contrast, in RunA, since the "parallel" and "perpendicular" directions are indistinguishable, we will only refer to the *longitudinal* and *perpendicular* directions, with respect to the filament.

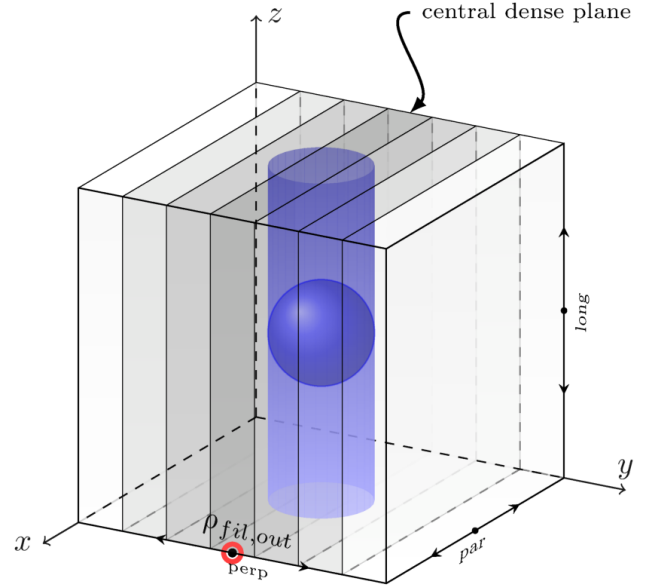

**Figure 1.** Illustration of the setup for the density field in the stratified simulation, RunS.

The various components (stratification, filament, and core) of the initial density field are set up as successive enhancements over a uniform density background with Gaussian profiles in one, two or three directions, as illustrated schematically in Fig. 1. The stratification (in RunS only) is set up as a Gaussian enhancement along the $y$-direction only, peaking at half the length of the box in this direction, with a standard deviation of 1.38 pc and an amplitude of 50% over the uniform background. The filamentary component is set up as a Gaussian enhancement in the $x$ and $y$ directions with a standard deviation of 0.69 pc and amplitude of 50% of the uniform background above its corresponding background (uniform cloud in RunA and the stratified background in RunS), centered at the middle point of the $(x,y)$ plane. Finally, the core was set up as a Gaussian enhancement in all three directions, centered at the middle of the numerical box, with a standard deviation of 0.50 pc, and an amplitude of 100% of the uniform background over the filamentary component. After all components are added up, the whole density field is renormalized in order to recover the mean density of $100\,\mathrm{cm}^{-3}$.

**Mathematically, the initial density in the $i$-th cell in the numerical grid is given by**

$$\rho(\boldsymbol{x}_i) = \frac{\rho'(\boldsymbol{x}_i)}{A}, \qquad (1)$$

**where**

$$\rho'(\boldsymbol{x}_i) = \rho_0 + \sum_{j=1}^{3} \rho_j \mathcal{G}_j(\boldsymbol{x}_i); \qquad (2)$$

$$\mathcal{G}_j(\boldsymbol{x}_i) = \exp\left[-\frac{(x_i^2 + y_i^2 + z_i^2)}{2\sigma_j^2}\right]; \qquad (3)$$

$\boldsymbol{x}_i = (x_i, y_i, z_i)$ **is the position vector of cell $i$ with respect to the center of the box; the values $j = 1, 2, 3$**

**Table 1.** Parameters of the initial density perturbations (cf. eq. (2)).

| $j$ | Perturbation component | $\mathcal{G}$ | $\rho_j$ [%] | $\sigma_j$ [pc] |
|---|---|---|---|---|
| 1 | Stratification | $\mathcal{G}(0, y_i, 0)$ | 50 | 1.38 |
| 2 | Filament | $\mathcal{G}(x_i, y_i, 0)$ | 50 | 0.69 |
| 3 | Core | $\mathcal{G}(x_i, y_i, z_i)$ | 100 | 0.5 |

**Table 2.** Selected snapshots used in several plots.

| Run | Snapshot | Time $[t_{\rm ff}]^1$ | [Myr] |
|---|---|---|---|
| RunA | 1 | 0.03 | 0.11 |
|      | 8 | 0.27 | 0.89 |
|      | 16 | 0.53 | 1.77 |
|      | 24 | 0.79 | 2.65 |
|      | 32 | 1.05 | 3.53 |
|      | 33 | 1.09 | 3.64 |
| RunS | 1 | 0.03 | 0.11 |
|      | 8 | 0.26 | 0.88 |
|      | 9 | 0.30 | 1.00 |
|      | 16 | 0.53 | 1.77 |
|      | 24 | 0.79 | 2.65 |
|      | 25 | 0.82 | 2.76 |
|      | 31 | 1.02 | 3.42 |
|      | 32 | 1.05 | 3.53 |

[1] Free-fall time for the mean density of the simulation.

respectively denote the cases of the stratified, filamentary, and spherical density enhancements;

$$A = \int_{V_{\rm box}} \rho'(\boldsymbol{x}) d^3 \boldsymbol{x}, \qquad (4)$$

is the full box density normalization; and the values of the $\rho_j$ and $\sigma_j$ are given in Table 1. Also, Fig. 2 shows the resulting initial density profiles along the various directions for the two simulations. For practical reference, the measured full width at half maximum (FWHM) **of each component is** also **indicated in the figure by the numbers next to each line**.

The gas is initially at rest, and no gravity-counteracting forces such as a magnetic field or small-scale turbulence are included, so gravitational contraction starts immediately. This setup represents the premise of the GHC scenario that substructures begin their local gravitational contraction when their masses become larger than the mean Jeans mass in their parent structure as a consequence of the large-scale contraction of the latter, which progressively reduces the global mean Jeans mass (Vázquez-Semadeni et al. 2019).

For the advanced stages of the evolution, to identify the boundary of **the filament**, we have chosen to use a density threshold criterion over the density of **its background medium. In RunS, the filament's background medium is the dense plane, while in RunA it is the whole numerical box. Thus the filament's boundary is defined by**

$$\rho_{\rm bnd, fil} = \alpha_{\rm fil}\, \rho_{\rm fil, out}, \qquad (5)$$

with $\alpha_{\rm fil} = 1.3$ and $\rho_{\rm fil, out}$ **being the density at the location indicated by the solid black dot with a surrounding red circle in Fig. 1. This point is chosen to represent the density of the background medium (the numerical box in RunA or the dense plane in RunS) far from the core.** The value of $\alpha_{\rm fil}$ was selected simply as a slight enhancement over the mean density of the background.

Additionally, it is useful to identify two different regions along the filament, depending on whether they contain the core or not for comparison with observed cores and filaments, respectively. The *off-core* region, roughly halfway between the center of the core and the border of the computational box, safely removes the central core while not being too strongly affected by periodic boundary effects. The other is the *core* region, which extends from the center of the **numerical box to the boundary of the filament on the plane perpendicular to its axis**.

To investigate the evolution of the filament-core system, we have chosen some selected snapshots in both runs (see Table 2).

## 3 RESULTS

In this section, we first describe the uniform-background, axisymmetric simulation (RunA), and then the stratified-background simulation (RunS). All figures are labeled with the type of background setup, the time in terms of the free-fall time, and the orientation along which we have calculated the plotted quantities for each panel.

### 3.1 Overall evolution

Since the initial conditions we use are globally Jeans-unstable in both runs, the simulations begin to undergo gravitational contraction as soon as they start evolving. Nevertheless, it is important to note that the contraction never becomes completely radial, but instead maintains the filamentary structure throughout the (prestellar) evolution, developing a filamentary collapse flow directed toward the central core. This is in agreement with the large-scale numerical simulations of GMC formation and evolution of Gómez & Vázquez-Semadeni (2014), in which the global and hierarchical gravitational contraction of the cloud generates filamentary structures onto which the rest of the cloud accretes.

Figure 3 shows cross sections at various times of the volume density field of both simulations on the planes passing through the center of the computational box (see labels), with the normalized velocity field overlaid. It is noteworthy that, due to the presence of the stratification, RunS produces a ribbon-like filament and, also the core is thinner in the direction perpendicular to the stratification. The horizontal lines define a slice across the filament at a fixed distance $L_{\rm J, init}/2 \approx 1.12\,{\rm pc}$ away from the boundary at the initial conditions, where we have computed various physical quantities (cf. Sec. 3.2).

The density field is almost uniform along the filament and away from the core, and in the background region away from the filament. Also, as suggested in the less idealized simulation of Gómez & Vázquez-Semadeni (2014), the velocity field smoothly changes direction as it approaches the filament, being mostly perpendicular to the filament at large distances from it, but becoming longitudinal in the filament's central axis, pointing towards the core. It is important to note that *no shocks develop during the prestellar evolution.* In addition, the velocity field direction remains quite constant throughout the evolution, suggesting an approximately stationary flow. We return to this point below in Sec. 3.2.

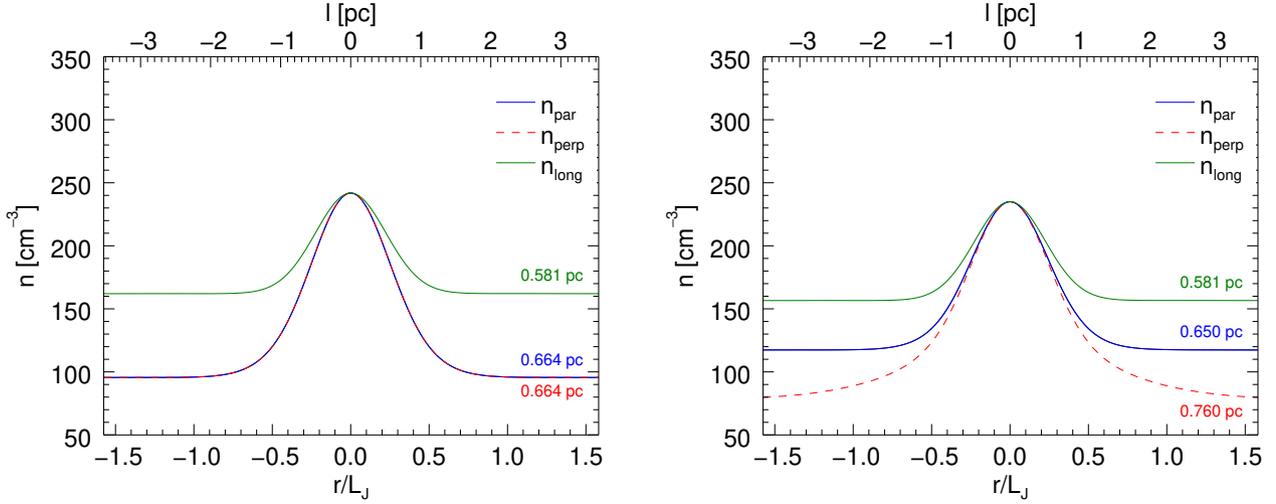

**Figure 2.** Density profiles at the initial conditions along the three main directions, passing through the center of the computational box for the axisymmetric simulation (RunA, left panel) and the stratified one (RunS, right panel). The red lines represent the direction perpendicular to the dense plane; the blue lines, represent the direction parallel to (on) the plane (perpendicular to the filament), and the green lines, the longitudinal direction (along the filament). The numbers on the right-hand side of the panels correspond to the FWHM of the density peak for each of the profiles. Note that, in RunA, the parallel and perpendicular profiles overlap due to the axial symmetry.

Figure 4 shows the density profiles for the selected snapshots in both simulations, with RunA shown in the two top panels and RunS shown in the bottom three panels, respectively. The various lines shown correspond to the selected snapshots, listed in Table 2. The longitudinal slices (right top and bottom panels) confirm that the density in the filament, away from the core, increases steadily in time through accretion from the cloud, although in RunA the increase rate appears to decrease towards the final stages of the prestellar evolution, suggesting an approach to stationarity.

Also, note that for RunS, the density at the dense plane increases, even far from the filament (bottom left panel, far from the core), while the density of the box away from the plane decreases (middle bottom panel, far from the core). Finally, the density in the filament also increases (bottom right panel, far from the core). This indicates that *there is accretion from the box onto the plane, from the plane onto the filament, and from the filament onto the core, thus constituting an extremely anisotropic and hierarchical accretion flow.*

Figure 5 shows the evolution of the linear mass density (often misleadingly referred to as the "line mass" in the literature) for the filament+core system (red lines) and the filament alone (blue lines). The green lines in the bottom panels show the evolution using values for the core and filament density thresholds from the initial conditions.

Interestingly, the filament evolution seems to transition from a regime of *increasing* rates of linear density growth to a regime of *decreasing* growth rates, again suggesting an approximation to a stationary regime, **both in the on-core and off-core positions**. This seems to be more **noticeable** in the stratified case. However, full stationarity (i.e., constant **linear density**) is clearly not fully reached during the prestellar phase investigated in this work.

Figure 6 shows the radial profiles for the number density, column density, the *radial* and *longitudinal* velocities for RunA (left column) and for RunS, both in the direction parallel to the dense plane (middle column) and perpendicular to it (right column). The solid lines show the profiles at the center of the box (*i.e.*, at the position of the core), while the dashed lines show the profiles midway between the core and the boundary; *i.e.*, on the off-core region of the filament. In this figure, we use a logarithmic radial axis to emphasize the internal structure of both the core and the filament.

From the solid lines, we can see that the collapse in the on-core position proceeds from the outside-in (that is, with the velocity peak removed roughly one Jeans length from the center, Whitworth & Summers 1985; Keto & Caselli 2010), in a similar way to the spherical case described in Paper I and other works (e.g., Whitworth & Summers 1985; Gómez et al. 2007; Gong & Ostriker 2009). Early in the evolution, at the on-core position (solid lines), the velocities are largest at large radial distances from the core's center, while the inner parts develop a velocity profile roughly linear with radius. At snapshot 9, $t \sim 0.3\, t_{\rm ff} \sim 1.0\,{\rm Myr}$, a transonic point appears $\sim 0.5\,{\rm pc}$ away from the center, that then splits into two points that move in opposite directions, *i.e.*, one outwards and one inwards. At later times, these transonic points enclose a region of almost uniform supersonic inward velocity. The density profile resembles that of a Bonnor-Ebert sphere, being flat in the inner region where the velocity is uniform, while approaching an $r^{-2}$ profile in the exterior parts.

Similarly, Fig. 7 shows the **longitudinal profiles of the volume density, column density, and longitudinal velocity along the central axis of the filament** in the last snapshot of each simulation at the on-core position. The behaviour is quite similar to that of the core on top of the uniform density background described in Paper I, with the main difference being that the flat inner part of the density and column density profiles for the core are more extended.

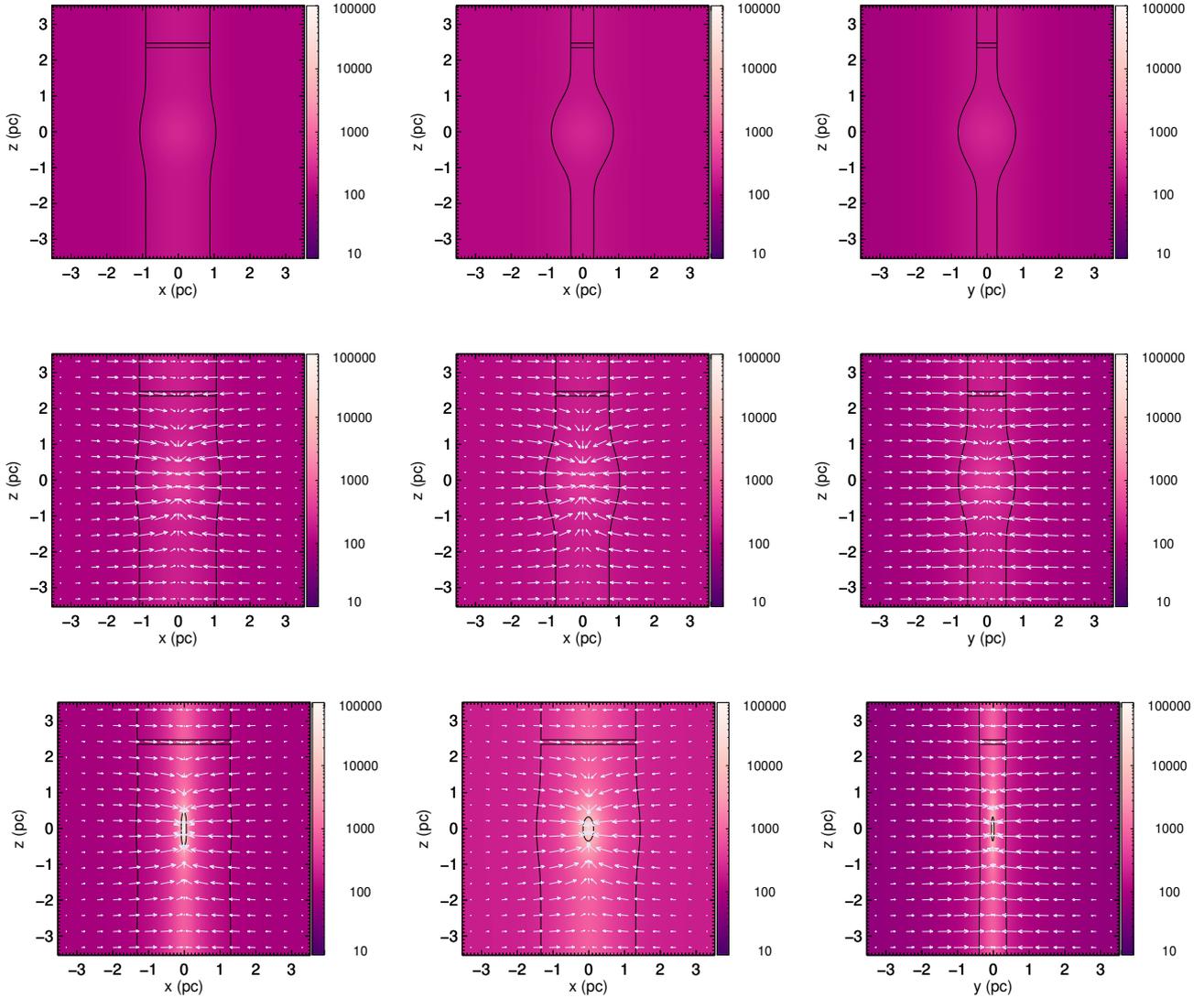

**Figure 3.** Density images for RunA on the $y = 0$ plane (left column) and RunS on the dense plane (middle column) and perpendicular to it (right column) at $t = 0.00 t_{\rm ff}$ (top panels), $t = 0.53 t_{\rm ff}$ (middle panels) and at $t = 1.02 t_{\rm ff}$ (bottom panels). Velocity vectors are normalized to the maximum velocity for each simulation. The evolutionary point in time of each panel corresponds to that in Table 1. The black lines in the upper part of each filament show the slices where the mass flux measurements are made in Sec. 3.2.

Figure 9 shows the radial column density profile of the filament **at the off-core position** and a **least-squares** fit of a Plummer-like function of the form

$$\Sigma_{\rm P}(r) = \frac{A_p \, \rho_c \, R_{\rm flat}}{\left[1 + \left(\frac{r}{R_{\rm flat}}\right)^2\right]^{\frac{p-1}{2}}} \quad (6)$$

at the end of the simulations. The parameters of the fit are the radius of the central flat part of the filament's radial profile, $R_{\rm flat}$, its power-law exponent $p$, and $A_p$ (see also Tables 3-4), which is a finite constant factor for $p > 1$. For comparison, it also shows the profile for an infinite, hydrostatic isothermal cylinder, for which $p = 4$ (Ostriker 1964)[2], and the derived profile for observed filaments, for which $p = 2$ (Arzoumanian et al. 2011). We can see that the fitted values of the slope ($p = 1.35$ for RunA and, $p = 1.25, 1.69$ for RunS, in the parallel and perpendicular directions, respectively), are in general less than the observed slope, **with the value for RunS in the direction perpendicular to the dense plane coming closest to the observed value. Moreover, in Fig. 8 we show the evolution of $p$ for the two runs, and note that it is not increasing towards the values typically observed in *Herschel* filaments (e.g., Arzoumanian et al. 2011) nor in self-consistent simula-** tions of filament formation (e.g., Gómez & Vázquez-Semadeni 2014). One possible reason for this may be that the periodic boundaries are much closer to the filament and core in our present simulations than in those of the latter authors. We plan to address this issue in a future study employing a different numerical scheme.

---

[2] For an infinite isothermal filament in hydrostatic equilibrium, $A_p = \pi/2$ and, $R_{\rm flat}$ corresponds to the thermal Jeans length at the center of the filament.

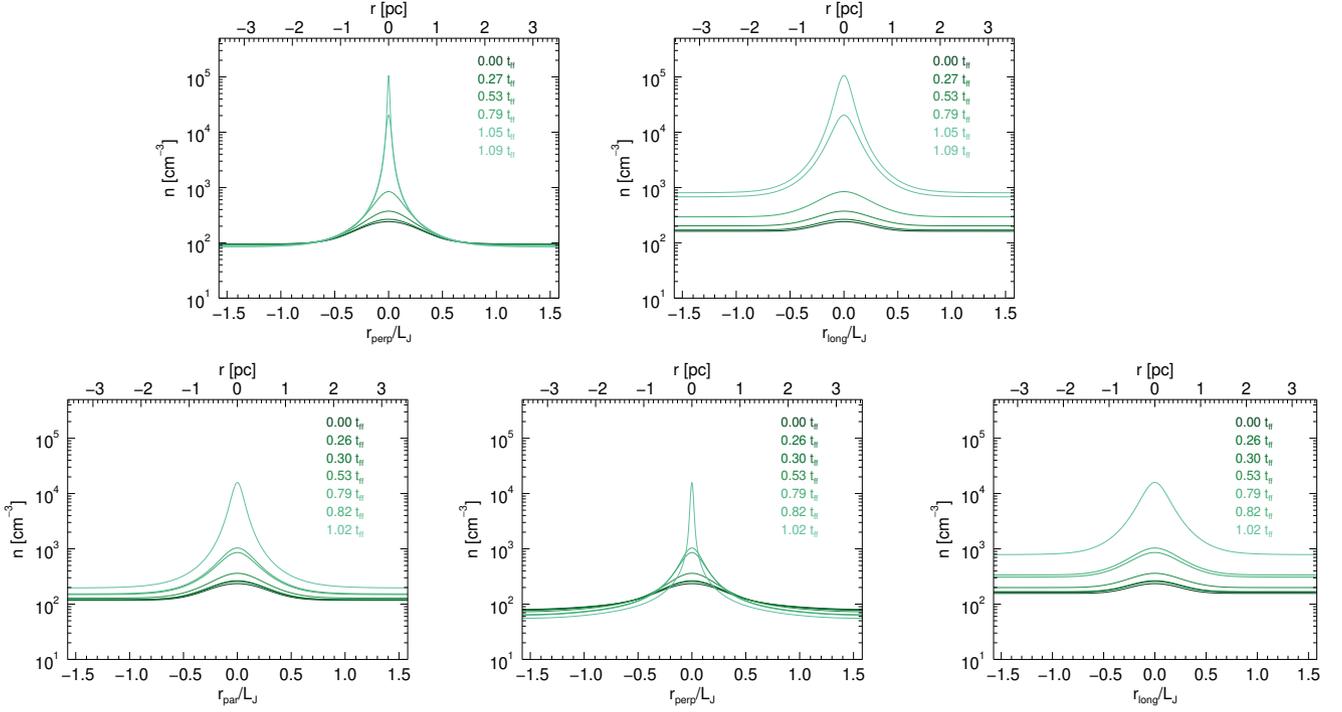

**Figure 4.** Evolution of the density profile for RunA perpendicular to the filament (top left panel) and along the filament (top right panel), and for RunS, perpendicular to the filament on the dense plane (bottom left panel), perpendicular to the filament and to the dense plane (bottom middle panel) and along the filament (bottom right panel). The various colored lines in all panels correspond to the snapshots listed in Table 2. **Note that the last curve shown for RunA corresponds to a slightly later time than the last curve for RunS (cf. Table 2, and therefore appears to reach a higher peak density.**

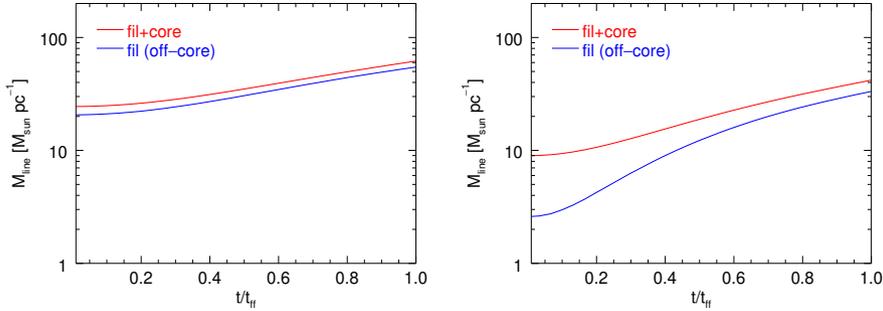

**Figure 5.** Evolution of the linear density (mass per unit length, $M_{\rm line}$) of the filaments, **at the on-core position (red lines) and the off-core position (blue lines)**. RunA is shown in the left panel, and RunS is shown on the right.

Figure 10 shows the evolution of $R_{\rm flat}$ obtained by fitting the column density profiles (see Figure 9) for both runs. As we can see from both panels, **at early times the fitted values of $R_{\rm flat}$ are quite larger than the typical values of $\sim 0.1$ pc reported from *Herschel* observations. On the other hand, at later times, this typical size scale** is approached, although the evolution of $R_{\rm flat}$ does not seem to approach it asymptotically, and instead seems to decrease in an accelerated fashion, so that it may eventually become smaller than the typical *Herschel* value. Also, it is noteworthy that the evolution of $R_{\rm flat}$ in the perpendicular direction in RunS is very similar to the evolution in RunA. In RunA, $R_{\rm flat}$ decreases by a factor of $\sim 4.5$ between the initial and final states, while for RunS it varies by a factor of $\sim 2.9$ and $\sim 9.9$ in the parallel and perpendicular directions, respectively.

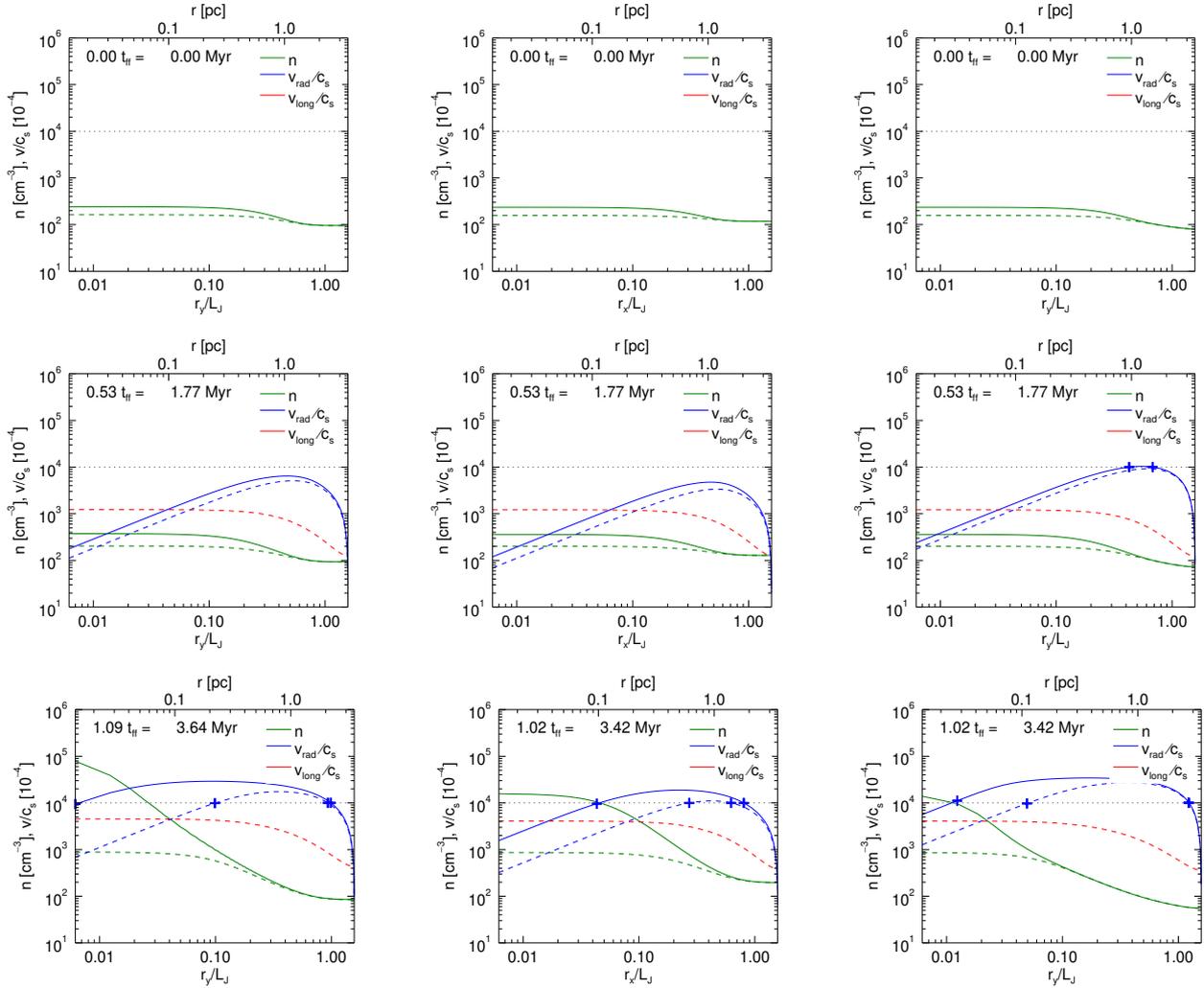

**Figure 6.** Radial profiles of the volume density and radial and longitudinal velocities measured at the on-core (solid lines) and off-core (dashed lines) positions for RunA (left column) and RunS, on the dense plane (center column) and perpendicular to it (right column), at $t = 0.00 t_{\rm ff}$ (top panels), $t = 0.53 t_{\rm ff}$ (middle panels) and $t \approx 1.02 t_{\rm ff}$ (bottom panels). Note that at the on-core position, the longitudinal (parallel to the filament) velocity is zero at all radii. **The black dotted horizontal line indicates the sound speed, and the "+" signs indicate the "sonic points", where the velocities reach the sound speed.**

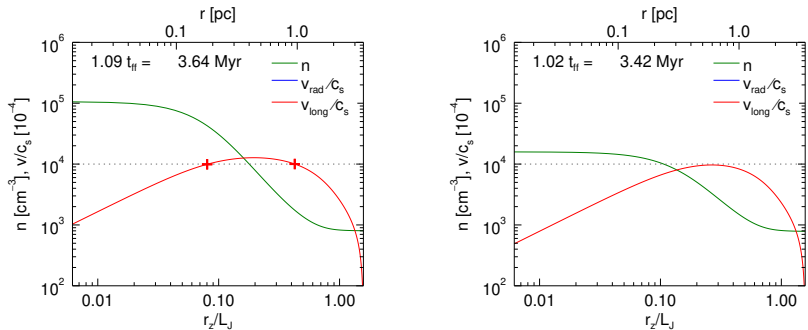

**Figure 7.** Similar to Fig. 6 but for the longitudinal profiles of the volume density and longitudinal velocity towards the end of the evolution along the central axis of the filament for RunA (left panel) and RunS (right panel). Note that the radial velocity vanishes at the central axis at all positions along the filament's length, **and so this curve is not shown. The black dotted horizontal line indicates the sound speed, and the "+" signs indicate the "sonic points", where the velocities reach the sound speed.**

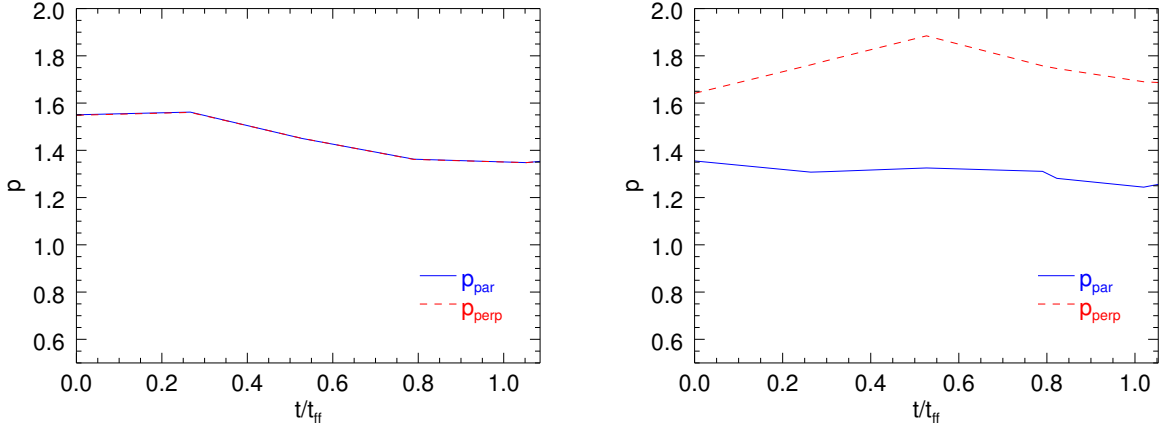

**Figure 8.** Evolution of the parameter *p* for RunA (left panel) and RunS (right panel).

**Table 3.** Fitted parameters for the column density profile for RunA.

| Snapshot | $R_{\rm flat}$ (pc) | $p$ | $A_p$ |
|---|---|---|---|
| 0  | $1.44 \pm 5.26$e-4 | $1.55 \pm 1.00$e-2 | $2.27 \pm 8.33$e-4 |
| 16 | $0.99 \pm 3.49$e-4 | $1.45 \pm 5.43$e-3 | $1.94 \pm 8.70$e-4 |
| 33 | $0.19 \pm 6.52$e-4 | $1.35 \pm 2.96$e-3 | $0.17 \pm 2.15$e-4 |

**Table 4.** Fitted parameters for the column density profile for RunS along parallel (top) and perpendicular (bottom) orientations.

| Snapshot | $R_{\rm flat}$ (pc) | $p$ | $A_p$ |
|---|---|---|---|
| 0  | $1.46 \pm 5.31$e-4 | $1.35 \pm 9.87$e-3 | $2.20 \pm 8.35$e-4 |
| 16 | $1.18 \pm 4.32$e-4 | $1.32 \pm 7.03$e-3 | $1.84 \pm 7.09$e-4 |
| 31 | $0.35 \pm 1.15$e-4 | $1.24 \pm 1.45$e-3 | $0.33 \pm 6.33$e-5 |
| 0  | $1.46 \pm 4.78$e-4 | $1.64 \pm 9.38$e-3 | $2.62 \pm 9.16$e-4 |
| 16 | $1.18 \pm 3.12$e-4 | $1.89 \pm 5.81$e-3 | $2.43 \pm 7.32$e-4 |
| 31 | $0.19 \pm 4.04$e-5 | $1.69 \pm 8.93$e-4 | $0.79 \pm 2.07$e-4 |

### 3.2 Mass flux in the filament and a possible approach to a stationary regime

As discussed in Sec. 3.1 (cf. Fig. 3), the velocity field in the simulations tends to remain constant in space, first in direction and then in magnitude as time advances, suggesting an approach to a stationary regime. To further search for evidence of this, in Figure 11 we plot **the relative differences between the final and initial ($t = 0.03 t_{\rm ff}$) histograms** of the angles between the velocity vectors and the horizontal axis from the panels of Fig. 3, normalized by the final histogram. In both runs, it is seen that the relative difference frequency distribution of these angles **varies by at most $\lesssim 10\%$, and much less than that in most cases**, supporting the view that the velocity field remains approximately stationary.

In fact, an approach to stationarity is expected. As can be seen from the bottom panels of Fig. 5 and from Fig. 6, both the filament's linear density $M_{\rm line}$ and longitudinal velocity $v_z$ increase monotonically over the entire prestellar evolutionary stage we simulate. **Moreover, the longitudinal profile of the longitudinal velocity component, shown in Fig. 7, is seen to be nearly flat from $r_z \sim 0.1 L_{\rm J}$ (or $\sim 0.3$ pc) to $r_z \sim 0.8 L_{\rm J}$ ($\sim 1$ pc). The drop near the boundary is an artifact of the boundary conditions. Moreover, in the simulations of Gómez & Vázquez-Semadeni (2014)**, in which the filament develops self-consistently within the cloud and far from the boundaries, the longitudinal velocity varies very mildly along the full length of the filament, and the filament's length remains stationary, because it accretes material from the cloud also at its tail. Therefore, the behavior of $v_z$ can be extrapolated to the full length of the filament. From all of these considerations, we can safely assume that the longitudinal mass accretion rate along the filament, $\dot{M} = M_{\rm line} v_z$, **should also increase with time**.

Now, it also appears safe to assume that the radial accretion rate from the cloud onto the filament, which feeds the filament, varies on longer timescales than the variation of the longitudinal accretion rate from the filament onto the core (which drains the filament), because the former evolves on the cloud free-fall time, which is longer because the cloud is at lower density. The increase in the filament's linear mass density suggests that the radial accretion is initially larger than the longitudinal one, causing the filament's linear density to continue growing. But then, the increase in the longitudinal accretion rate suggests that it should approach the radial rate.

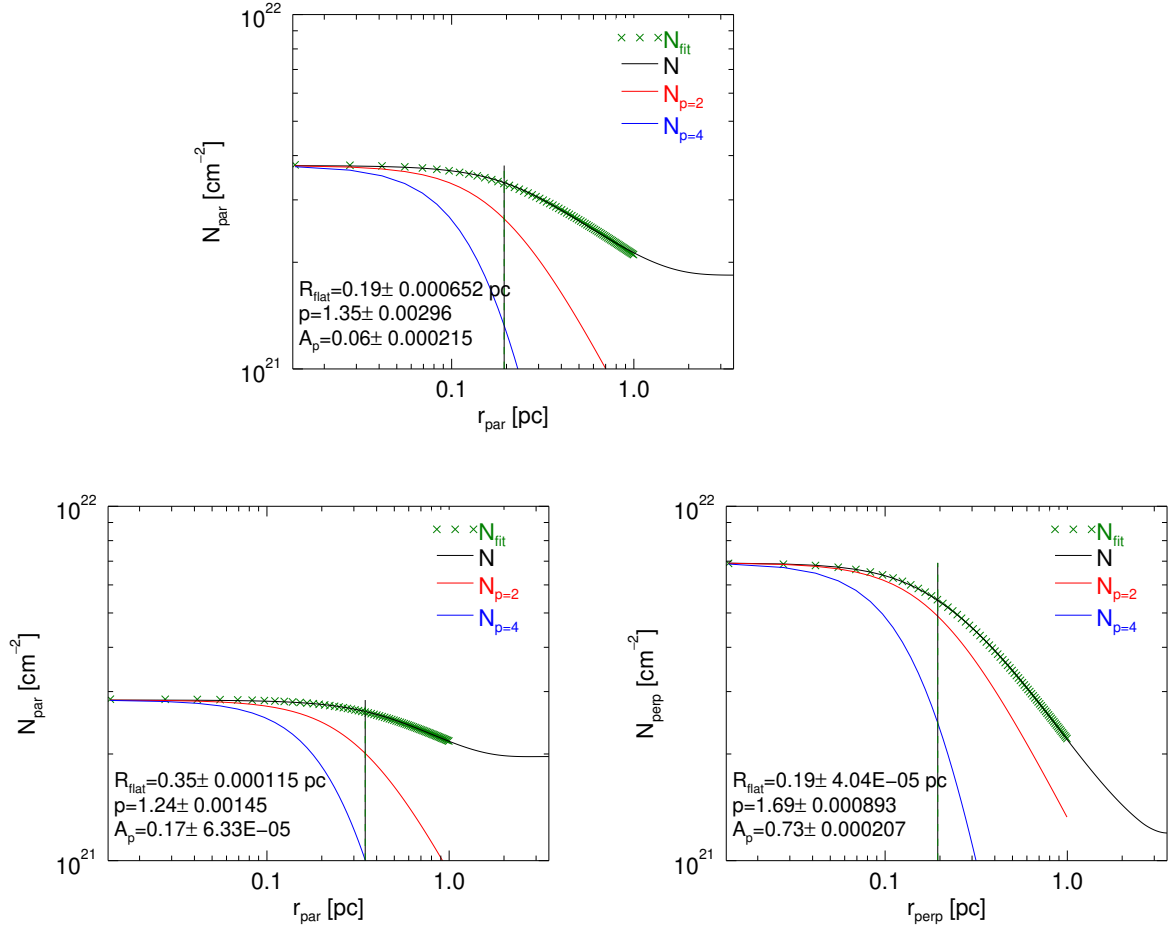

**Figure 9.** Radial column density profiles of the filament at the off-core position (black) close to the end of the simulations (RunA, top panel; RunS, bottom panels). The green crosses indicate the points used for computing the best-fit parameters using a Plummer-like function. Also shown are Plummer profiles, given by eq.( 6), with $p = 4$ (adequate for an infinite hydrostatic isothermal cylinder; Ostriker 1964), and with $p = 2$ (adequate for the filament sample of Arzoumanian et al. 2011). **The vertical lines indicate the radial position of $R_{\rm flat}$ in each case.**

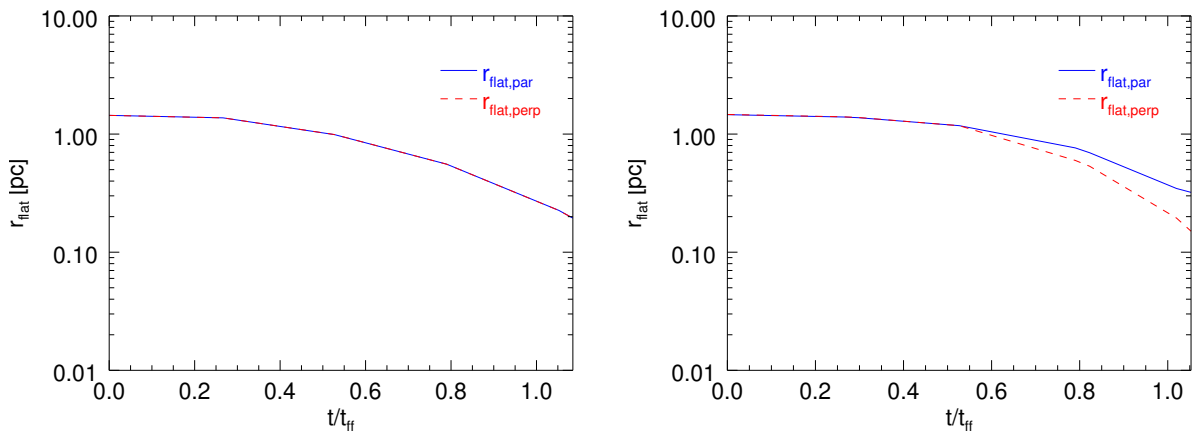

**Figure 10.** Evolution of $R_{\rm flat}$ for RunA (left panel) and RunS (right panel).

Conversely, assume that eventually the longitudinal rate becomes larger than the radial one. In this case, the filament would lose mass and decrease its linear density, decreasing the longitudinal accretion rate. Therefore, it appears that the longitudinal rate will tend to always balance the radial rate, as long as the variation in the latter occurs over significantly longer timescales than that of the former.

To test numerically the hypothesis of a stationary regime, we have selected a slice of the filament of thickness 10 pixels, at a distance $l = L_{\rm J,init}/2$ from the boundary, to measure the mass per unit length (the linear density), as well as the longitudinal and radial mass fluxes in the filament. This choice for the slice was made so that the slice is sufficiently removed both from the core, since we are interested in the filament in this case, and from the boundary of the numerical box, where the longitudinal flow velocity is forced to remain at zero. These chosen slices for each case are shown by the black lines in Fig. 3.

Consider Fig. 12. If the flow approaches stationarity, then the mass flux *into* the slice, $\dot{M}_{\rm in}$ (both through the top "lid" of the slice, in the direction toward the core, as well as in the radial direction, due to accretion from the cloud) must approach the flux *out of* the slice, $\dot{M}_{\rm out}$, through its bottom lid. That is, we expect that the total mass change rate ratio,

$$\mu_{\rm tot} \equiv \frac{\dot{M}_{\rm in}}{\dot{M}_{\rm out}} \sim 1, \qquad (7)$$

where

$$\dot{M}_{\rm in} = \int_A \rho \boldsymbol{v} \cdot d\boldsymbol{A} + \int_B \rho \boldsymbol{v} \cdot d\boldsymbol{B},$$

$$\dot{M}_{\rm out} = \int_C \rho \boldsymbol{v} \cdot d\boldsymbol{C}, \qquad (8)$$

$\boldsymbol{v}$ is the velocity vector, $\boldsymbol{A}$ is the area of the top lid, $\boldsymbol{B}$ is the perimetral area of the slice, and $\boldsymbol{C}$ is the area of the bottom lid.

Figure 13 shows the evolution of $\mu_{\rm tot}$ through the slice for the two runs. As we can see, the mass flux ratio for **both runs** has not yet reached unity by the time they stop, although $\mu_{\rm tot}$ is **still clearly decreasing** and may eventually reach unity. **This suggests that, if the stationary regime is reached, it should happen during the protostellar stage in the central core.**

Another relevant diagnostic is the ratio of the mass accretion rate onto the slice through the perimetral area, $\dot{M}_{\rm in,B}$, to the total mass rate of change of the **whole** slice, $\dot{M}_{\rm slice}$. **These rates are respectively** given by

$$\dot{M}_{\rm in,B} = \int_B \rho \boldsymbol{v} \cdot d\boldsymbol{B}. \qquad (9)$$

and

$$\dot{M}_{\rm slice} = \dot{M}_{\rm in} - \dot{M}_{\rm out}. \qquad (10)$$

We then expect that, at early times, most of the mass change in the slice is due to the perimetral accretion, and so the mass rate ratio

$$\mu_{\rm B} \equiv \frac{\dot{M}_{\rm in,B}}{\dot{M}_{\rm slice}} \qquad (11)$$

should be close to unity. On the other hand, at late times, if the mass of the slice tends to a constant (*i.e.*, $\dot{M}_{\rm slice} \to 0$), then this ratio should **increase**, indicating that the mass accreted through the perimetral area plus the mass accreted through the top lid, is **fully** expelled through the bottom lid, **so that** the perimetral accretion no longer contributes to an increase in the slice's mass.

Figure 14 shows the evolution of $\mu_{\rm B}$ for the two simulations. For both, this ratio is indeed $\gtrsim 1$ at early stages. This means that the mass flux through the perimetral area is close to, but slightly larger than, the total mass increase rate of the slice. This in turn implies that, during the early stages, this perimetral mass flux is the main driver of the mass growth of the slice, although some of it is lost by evacuation from the bottom lid.

On the other hand, at later times, the ratio begins to increase **for both runs**, indicating that the evacuation (through the bottom lid) increases in relation to the perimetral inflow rate, so that the mass of the slice begins to approach constancy (its mass rate of change decreases); i.e., the slice begins to approach stationarity. Figure 15 shows the evolution of the mass of the slice (blue lines) for the two runs, **which is seen to grow monotonically in both cases. However, it is seen that the rate of mass increase is first superlinear, and then seems to stabilize at $t \sim 0.4$-$0.5 t_{\rm ff}$, when the growth of the slice's mass becomes approximately linear with time. This approximately coincides with the time at which $\mu_{\rm B}$ begins to increase. We cannot determine, however, whether the growth of the slice's mass will eventually saturate in the protostellar stage.**

## 4 SUMMARY AND DISCUSSION

### 4.1 Summary

The results from the previous section can be summarized as follows:

(i) We have numerically simulated the prestellar collapse of a filamentary perturbation containing a spherical central enhancement (a core) at its center. We considered two variants of this filamentary collapse, one with axial symmetry (RunA), and one with additional stratification in one of the directions perpendicular to the filament (RunS). The latter represents the case in which the filament is in turn embedded in a flattened cloud.

(ii) The presence of the filamentary perturbation changes the symmetry of the collapse flow which, away from the core, proceeds first toward the filament, and smoothly changes direction as it approaches the filament axis, to becoming longitudinal there, *i.e.*, oriented towards the core. No shocks develop in the filament, in spite of the flow eventually becoming supersonic, in analogy with the prestellar collapse of a purely spherical core.

(iii) **To measure properties of the filament, and intending to represent common observational proce-**

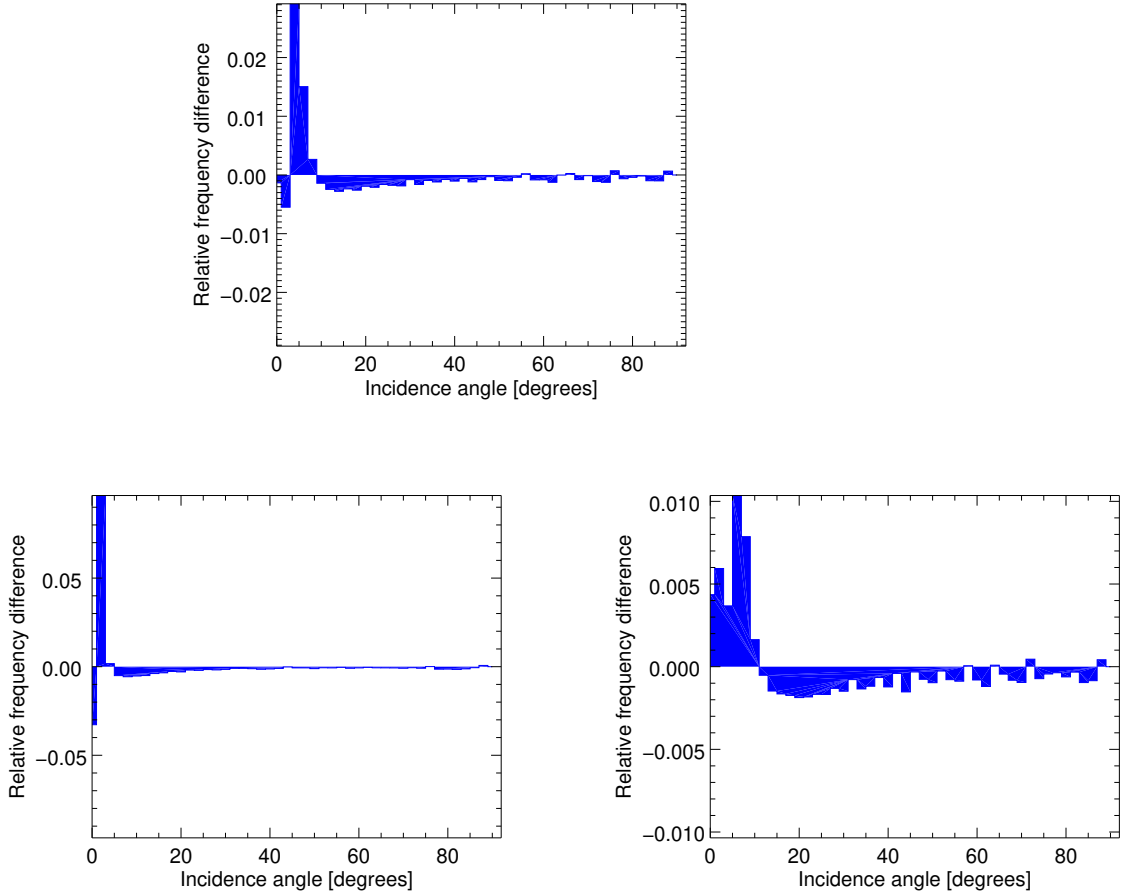

**Figure 11.** Relative difference between the final and initial histograms of the angle between the velocity vector and the horizontal axis in the plots of **Fig. 3** for RunA (top panel, corresponding to the left column in Fig. 3) and RunS, both on the central dense plane (bottom left panel, corresponding to the middle column in Fig. 3) and perpendicular to it (bottom right panel, corresponding to the right column of Fig. 3). It can be seen that the **relative difference is at most $\lesssim 10\%$, and much smaller than that in most cases, so that the orientation of the flow does not vary significantly throughout the evolution.**

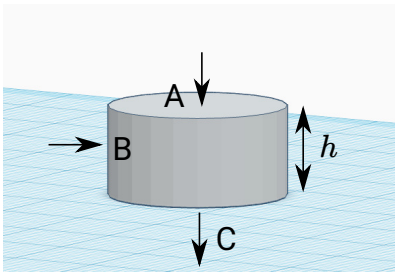

**Figure 12.** Illustration of mass flux through the filament's slice.

dures, we defined its boundary in terms of density enhancements above its parent structure (the full numerical box in RunA and the dense plane in RunS) as given by eq. (5). Thus defined, the filament evolves, growing in mass **and central density. However, the density increases much faster at the location of the central core than far from it.**

(iv) The filament is never hydrostatic, and instead **is always in the process of funneling** the material accreted through the boundary to the core. We suggested that **this accretion flow may tend towards a stationary state, which could constitute an attractor for the system. This is** because, if the longitudinal "drainage" of material is lower than the peripheral accretion, the filament's linear density must increase, increasing the longitudinal flux. Conversely, if the longitudinal flux exceeds the peripheral accretion, then the filament's linear density must decrease, reducing the longitudinal flux.

(v) In **order to determine whether the flow is approaching stationarity, in both runs we measured the change in the orientation of the velocity field from the early to the late stages of its evolution, as well as the mass fluxes into and out of a slice of the filament, across its perimetral and outer and inner "lid" surfaces. We found that the velocity field direction does not change significantly throughout the evolution, supporting the approach to stationarity. However, the mass fluxes across the slice cause a sustained mass increase in the slice, implying that stationarity** has not been reached. Therefore, we conclude that **stationarity is not reached during the prestellar evolutionary stage, and it remains to be investigated whether it may be reached during the subsequent protostellar stages of evolution. We plan to address this possibility in future work.**

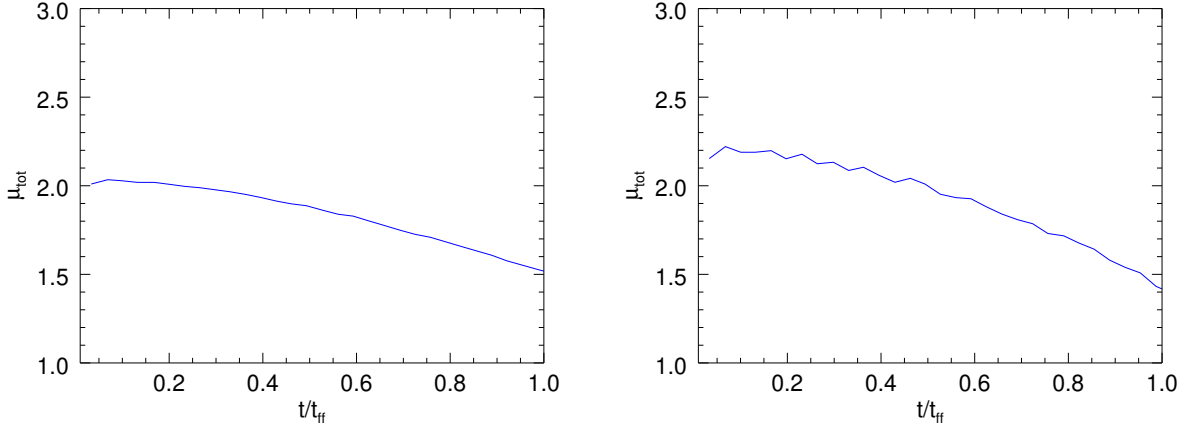

**Figure 13.** Evolution of the mass flux ratio, $\mu_{\rm tot}$, (eq.[7]) in the slice for Run A (left panel) and RunS (right panel).

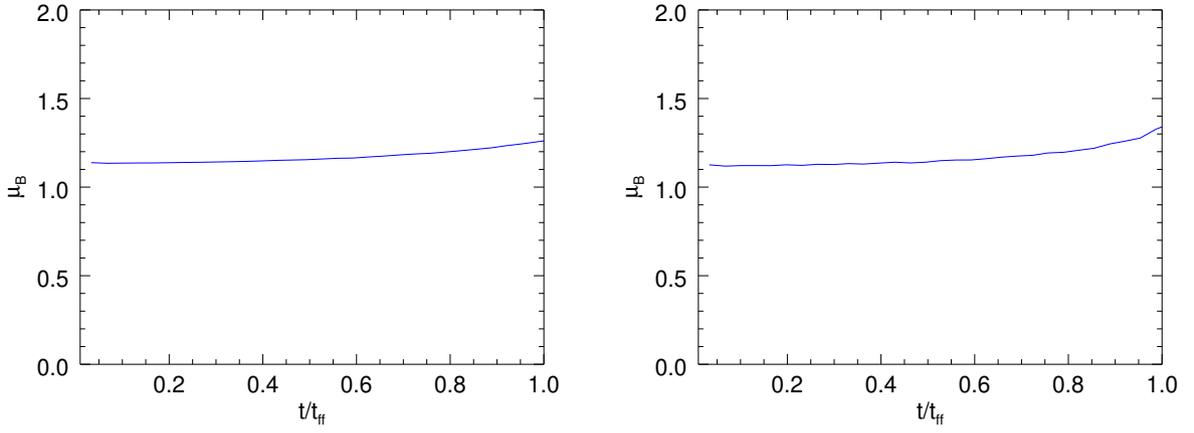

**Figure 14.** Evolution of the ratio, $\mu_{\rm B}$, of perimetral accretion rate to total mass change rate (eq.[11]) for the filament slice for RunA (left panel) and RunS (right panel).

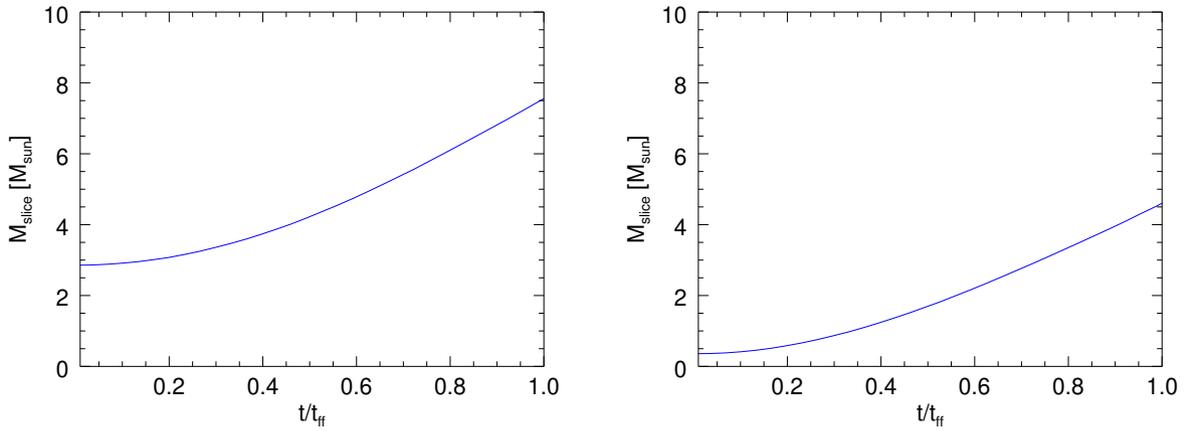

**Figure 15.** Slice mass *vs.* time for RunA (left panel) and RunS (right panel).

(vi) We found that the radial column density profile in the filament away from the core can be fitted by a Plummer profile, and that the radius of the flat central part slowly decreases in time, approaching the "typical" observed values of order 0.1 pc at the end of the simulations, near the time of protostar formation in the core. The profile on the filament retains its flattened shape near the filament axis even at the time when the central density in the core diverges (the time of protostar formation). This behaviour is due to the fact that, while the mass is accumulated at the core, it just "traverses" the filament but does not accumulate there, thus never causing a divergence of the density in the filament axis. Therefore filaments simply act as intermediary "funnels" for the material to flow from the cloud to the core.

(vii) **Finally, it is important to note that the filament-core system that develops in our simulations is the result of the self-gravity** *of the whole cloud*, **not just that of the filament itself. In fact, the filament-core system** *forms* **from the collapse of the cloud, and at the beginning its own self-gravity must be negligible, similarly to the case of the centermost part of a collapsing spherical core, which is Jeans-stable, and is being crunched by the infall of the rest of the core** (Gómez et al. 2021). **The system can be thought of as the tip of the iceberg of a larger-scale collapsing object.**

**4.2 Comparison to previous work**

Our results are consistent with **the properties of filaments developing self-consistently in** simulations of the formation and evolution of turbulent molecular clouds (e.g., Heitsch et al. 2009b; Smith et al. 2011; Gómez & Vázquez-Semadeni 2014), and reinforce the **suggestion** that the filaments constitute intermediate stages of the collapse. Our idealized setup, of a perfectly cylindrical **perturbation** with an **additional** spherical central **one**, allows us to focus on the essential flow features of filament-core systems **formed by anisotropic gravitational contraction**.

We find that, as in calculations of spherical collapse (e.g., Larson 1969; Penston 1969, see also Paper I), a shock does not develop anywhere in the system before the formation of a singularity (the time of protostar formation).

**Although we did not find conclusive evidence of the approach to stationarity, if such a steady state develops after the formation of a singularity (protostar), it may** be considered the filamentary analogue of the similarity flow developing in spherical structures after the formation of the singularity in flows that do not start from a singular hydrostatic initial condition (the singular isothermal sphere, or SIS), but rather have been growing from before the formation of the singularity (Whitworth & Summers 1985). For these, Murray et al. (2017) recently showed that the density at a given radius approaches a constant, implying that the flow becomes stationary. In this spherical case, the mass is drained into the singularity (the central protostar), so that the latter increases its mass, but the gaseous core may maintain a stationary density and velocity configuration (neglecting the increase in the core mass). In the filament, our results suggest that the drainage into the core may allow the filament to also develop stationarity, with our simulations **possibly approaching, although certainly not reaching** it. Moreover, in the case of our filament, because the center of the collapse flow (the center of the core) is far from the middle position on the filament in which we have measured the mass fluxes, the density at the filament axis remains finite, and allows the radial density and column density profiles to remain flat for a long time.

**The absence of a shock at the filaments' axis is a fundamental signature of the self-gravity-driven filamentary flow regime we have investigated, which proceeds smoothly from subsonic to supersonic velocities, much like the analytical solutions of spherical accretion onto a point mass (e.g.,** Bondi 1952) **and of spherical collapse during the prestellar stage (e.g.,** Whitworth & Summers 1985). **The "failure" to develop a shock at the filament's axis in our funnel-like flow,**[3] **the flattening of the radial density and column density profiles near the axis, and the decrease of the radial velocity towards the axis (cf. Figs.** 6–9) **suggest that the filament remains in a state similar to the prestellar stage of spherical core collapse until the time of the formation of a singularity at the core. The spherical prestellar evolution shares all of these features (**Whitworth & Summers 1985), **when the sphere's center is replaced by the filament's axis. The reason for this retardation appears to be the drainage of the material from the filament by the longitudinal flow towards the hub, halting or at least severely slowing down the growth of the density at the filament's axis as the stationary regime is approached.**

**The radial and longitudinal flows appear to develop roughly simultaneously** *while the filament grows*, **and so, following the development of the density and velocity field from sufficiently early times to allow them to co-evolve appears to be an essential requirement to obtain this regime. Indeed, for example, in simulations where the filament is set up at the initial conditions so that the cloud-to-filament and the filament-to-core accretions are not self-consistent, the filament can collapse radially or longitudinally (e.g.,** Seifried & Walch 2015), **because it is not adequately replenished by its environment, and/or produce** *strong* **accretion shocks at its boundary (e.g.,** Clarke et al. 2016; Heigl et al. 2020). **This is not observed in simulations of cloud formation and evolution in which the filaments arise and evolve self-consistently within the cloud (e.g.,** Heitsch et al. 2009b; Smith et al. 2011, 2014; Gómez & Vázquez-Semadeni 2014; Fogerty et al. 2017; Vázquez-Semadeni et al. 2017; Zamora-Avilés et al. 2017, 2019).

**The periodic boundary conditions in our simulation force the filament to remain uncollapsed (al-**

---

[3] Similar to the "conveyor belt" flow described by Longmore et al. (2014).

beit for a numerical condition rather than from end-on accretion from the cloud; see Sec. 4.3.2 and Heitsch 2013a), while the choice of initial conditions as moderate-amplitude perturbations—rather than setting up a fully-developed filament from the start—allows it to develop a self-consistent flow regime similar to that observed in the large-scale simulations.

In the absence of the longitudinal flow, the material would necessarily accumulate in the filament, and eventually a shock would develop at the filament's axis, possibly generating turbulence that provides additional support for the filament in the radial direction (e.g., Hennebelle & André 2013; Heitsch 2013b; Heigl et al. 2018, 2020).

Finally, it is important to note that, since real molecular clouds are turbulent, the turbulent flow may compete with the gravity-driven flow. If the turbulence contains more energy than the self-gravity, then it may disrupt the flow altogether. However, if self-gravity dominates, then the turbulent fluctuations will ride on top of the gravity-driven flow, and may produce *weak* shocks, either in the interior of the filament, possibly producing secondary structures such as "fibers" (e.g., Hacar et al. 2013), or in the periphery, producing mild accretion shocks (e.g., Bonne et al. 2020).

### 4.3 Caveats

#### 4.3.1 Applicability of the gravity-driven filamentary flow regime

It is important to emphasize that our study belongs to a different class of models in comparison with several other recent studies mentioned in Sec. 1, as it does not aim to produce a highly realistic model of a molecular cloud filament in the presence of multiple physical ingredients, like turbulence and magnetic fields. Instead, it aims to investigate the fundamental underlying flow regime driven by self-gravity in the presence of a filamentary plus a spherical perturbation, representing a hub/filament system. Under the GHC scenario, this may be the main driver of the formation and evolution of filaments in strongly self-gravitating molecular clouds. However, this means that our study most likely does not apply to the filamentary structures in non-self-gravitating gas, such as those in the diffuse cold atomic clouds.

Then again, it is not clear that some of the more diffuse molecular clouds are strongly self-gravitating, and the applicability of the gravity-driven regime to filaments in this type of clouds is in doubt. A typical example is the Polaris molecular cloud, whose filaments have been determined to have linear mass densities lower than the critical value for gravitational instability, and therefore are considered to be unable to form stars (e.g., André et al. 2010). However, it is important to note that, under the GHC scenario, molecular clouds are expected to accrete from their atomic envelopes, increasing their mass and star-formation activity over the period of a few million years. For example, Vázquez-Semadeni et al. (2018) have shown that the dense mass fractions and the star formation rates of the Lada et al. (2010) molecular cloud sample are consistent with the expectations from the evolutionary model of Zamora-Avilés & Vázquez-Semadeni (2014), while Camacho et al. (2020) have shown that a cloud evolving in the turbulent warm atomic medium transits from a Pipe-like stage to an infrared-dark-cloud-stage over the course of a few million years. Therefore, it is unclear whether the Polaris cloud is doomed to remain in its low-mass quiescent state, or whether it will able to evolve into an at least moderately active star forming region. So, the self-gravity-driven flow regime we have investigated may still be applicable to the filaments in the Polaris cloud and other similar clouds. Further research on the total mass reservoir available to these clouds is needed to define this issue.

#### 4.3.2 Limitations of the simulations

Because our numerical simulations have been performed with periodic boundary conditions, even for the self-gravity, the infall velocities are artificially set to zero at the boundaries, preventing accretion onto the numerical box, and reducing the value of the infall speed in the vicinity of the boundaries. Therefore, our simulations should be considered to provide a lower limit to the values of the infall speeds.

Also, since the infall speed is fixed at zero at the boundaries, the filaments cannot contract longitudinally. However, this is actually an advantage of the setup, since in larger-scale simulations where the filaments form self-consistently within a much larger cloud, they are also found *not* to contract longitudinally, because they are constantly being replenished by accretion from the cloud even end-on (e.g., Heitsch 2013a; Gómez & Vázquez-Semadeni 2014). Therefore, our more local simulations reproduce this feature of the larger, self-consistent ones, albeit for a numerical, rather than physical, reason. It is because of this that in Sec. 3.2 we have chosen to compare the radial and longitudinal mass fluxes at a sufficient separation from the boundaries, in order to minimize the boundary effects.

### ACKNOWLEDGEMENTS

We thank an anonymous referee for comments that have greatly improved the focus of the paper and helped us identify one glitch in the filament definition. R.N.-R. acknowledges financial support from PAPIIT grant IA103517 from DGAPA-UNAM to Bernardo Cervantes-Sodi. E.V.-S acknowledges CONACYT grant 255295.

### DATA AVAILABILITY

The data underlying this article will be shared on reasonable request to the authors.

This paper has been typeset from a TeX/LaTeX file prepared by the author.